\documentclass[pre,superscriptaddress,showkeys,showpacs,twocolumn]{revtex4-1}
\usepackage{graphicx}
\usepackage{latexsym}
\usepackage{amsmath}
\begin {document}
\title {Agreement dynamics on directed random graphs}
\author{Adam Lipowski}
\affiliation{Faculty of Physics, Adam Mickiewicz University, Pozna\'{n}, Poland}
\author{Dorota Lipowska}
\affiliation{Faculty of Modern Languages and Literature, Adam Mickiewicz University, Pozna\'{n}, Poland}
\author{Ant\'onio  Luis Ferreira}
\affiliation{Departamento de F\'{i}sica, I3N, Universidade de Aveiro,  Portugal}
\begin {abstract}
We examine some agreement-dynamics models that are placed on directed random graphs. In such systems a fraction of sites $\exp(-z)$, where $z$ is the average degree, becomes permanently fixed or flickering. In the Voter model, which has no surface tension, such zealots or flickers freely spread their opinions and that makes the system disordered. For models with a surface tension, like the Ising model or the Naming Game model, their role is limited and such systems are ordered at large~$z$. However, when $z$ decreases, the density of zealots or flickers increases, and below a certain threshold ($z\sim 1.9-2.0$) the system becomes disordered.  On undirected random graphs agreement dynamics is much different and ordering appears as soon the graph is above the percolation threshold  at $z=1$.
\end{abstract}

\maketitle

\section{Introduction}

Processes that tend to align neighbouring spins or make people to have the same opinion or share the same convention are examples of an agreement dynamics.
Although this dynamics is typically local, sometimes it has far-reaching consequences such as formation of a macroscopic ferromagnetic ordering, an overhelming support for a certain political party or an emergence of a common language. A spontaneous appearance of such patterns characterizes many complex systems and draws attention of scientists of multiple disciplines~\cite{loreto,durlauf,nowak}.  

Various statistical-mechanics models that aim to describe an agreement dynamics were examined; some representative examples include the Ising model~\cite{huang,stauffer}, the Naming Game model~\cite{steels,baronchelli-sharp} and the Voter model~\cite{liggett,fernandez}.  The behaviour of these models depends, of course, on the network of interactions between their building elements, i.e., spins, linguistic agents or voters.  
The networks that are simplest for numerical or analytical study  include various low-dimensional regular lattices or complete graphs, but in the context of complex-systems modelling, heterogeneous networks seem to be more relevant. These networks include, for example, scale-free networks, small worlds and random graphs~\cite{albert}, and various agreement-dynamics models have already been examined on such networks~\cite{mendesreview}. 

Another important ingredient, which should be taken into account while modelling complex systems, is a directedness of links. This feature reflects the fact that very often relations like ''A mimicks B'' or ''A hears B'' are not symmetric. However, the effect  of the directedness on the agreement dynamics has been not well understood yet. Some research in this field is related with conservation laws in Voter-like models \cite{serrano}, the role of long-range links on opinion formation \cite{jiang} or synchronization in Hodgkin-Huxley systems \cite{park}. Recently, we have examined the Ising model on directed random graphs and we have shown that for a ferromagnetic ordering to exist, a network must be sufficiently dense~\cite{liplip2015}. This ferromagnetic threshold value is considerably above the percolation threshold, which means that the Ising model on a directed random graph behaves qualitatively differently from the undirected version, where the ferromagnetic threshold coincides with the percolative one~\cite{dorogovtsev,leone}. Moreover, on the directed network the zero-temperature coarsening leads to the ferromagnetic state, which also differs from the undirected networks, where the model gets stucked in a certain disordered state~\cite{castellano,svenson,haggstrom}. Since the Ising model is one of the basic models of statistical mechanics, it would certainly be highly desirable to have a better understanding of such properties. Moreover, and that was the main motivation of our research, it would be interesting to confront the behaviour of Ising model with some  other agreement dynamics systems.  

\begin{figure}
\includegraphics[width=\columnwidth]{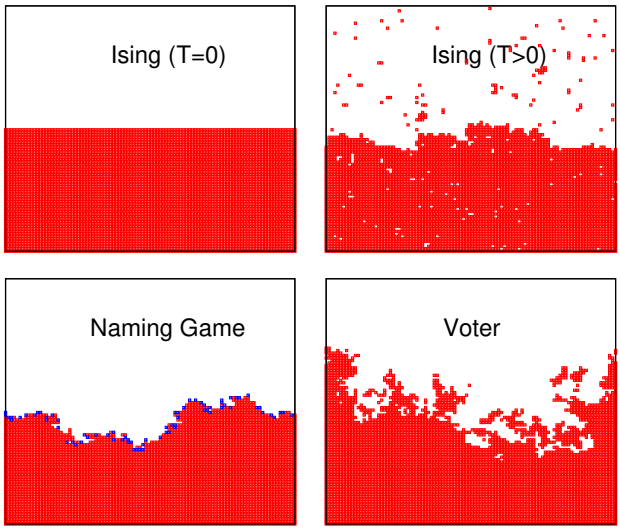}
\caption{Configuration snapshots of square-lattice (100x100) versions of models examined in the present paper. All models start from the configuration with the upper half  in one state and the lower half in the other. (i) The absence of bulk and interfacial noise for the Ising model at zero temperature implies that the initial configuration is actually a frozen state. (ii) For the Ising model at a positive temperature, the bulk noise can create some bubbles anywhere in the system but the surface tension limits the length of the interface. (iii) In the Naming Game, the interface might fluctuate (with an effective surface tension) but there are no bubbles far from the interface (a small fraction of sites at the interface, shown in blue, have two words in their inventories). (iv) In the Voter model, there is no bulk noise and the interface has no surface tension.}
\label{config}
\end{figure}

In the present paper, we examine two other agreement-dynamics models on directed (Poissonian) random graphs, namely the Naming Game and the Voter model. These models are qualitatively different from the Ising model (Fig.~\ref{config}).
There is no bulk noise in the Naming Game model  (similarly to the zero-temperature Ising model), but there is an interfacial noise (similarly to the Ising model at a positive temperature) as well as an effective surface tension~\cite{topology,dallasta}. 
Such a effective surface tension might be defined for finite-dimensional models and its usage in the context of random graph models should be however taken with some care. 
The Voter model~\cite{dornic} differs from the Naming Game mainly in the absence of a surface tension. An interesting feature of agreement-dynamics models on directed networks is the presence of either zealots or flickers, i.e., the sites that remain permanently fixed or constantly change their states. 
Such sites are known to considerably influence the agreement dynamics \cite{szymanski,waagen}.
In the Voter model, the absence of a surface tension implies that the zealots freely spread their opinions, which makes the system disordered.
Both for the Naming Game and the Ising model, the surface tension limits the influence of zealots and flickers, and this is why the system may order. However, when the number of links in the graph decreases, the fraction of zealots or flickers increases and at a 
certain point they destroy the long-range order in the system. Our results thus show that the surface tension, the type of noise generated by dynamical rules, and the directedness of the network considerably influence agreement-dynamics models.

\section{Naming Game}

\begin{figure}
\includegraphics[width=6cm]{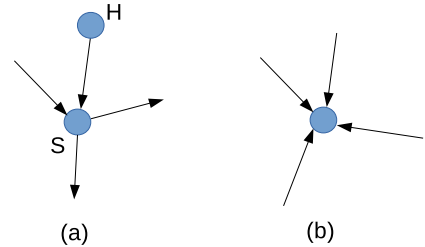}
\caption{(a) The randomly selected Speaker (S) selects the Hearer (H) from among sites that are in-connected with the Speaker. (b) In Naming Game, a site that has no out-links becomes a zealot. Such a site cannot be selected as a Hearer and its inventory remains permanently fixed.}
\label{graph}
\end{figure}

First, we analyse the Naming Game, which is a model that describes a process of reaching a consensus in a population of communicating agents~\cite{baronchelli-sharp}.
In our model, we have  $N$~agents placed at sites of a directed random graph of the average in-degree~$z$ (that is, of course, equal to the average out-degree). The agents try to negotiate a shared name for a given object. Each agent has its own inventory, which is a dynamically modified list of words (typically, with a random item at the beginning). The following act of communication between agents constitutes
an elementary step of the dynamics:
\begin{itemize}
\item The Speaker is  randomly selected. Then the Hearer is selected as one of the sites connected with the Speaker via its in-coming links (Fig.~\ref{graph}a). 
\item The Speaker selects a word randomly from its inventory and transmits it to the Hearer.
\item If the Hearer has the transmitted word in its inventory, the interaction is a success and both players maintain only the transmitted word in their inventories.
\item If the Hearer does not have the transmitted word in its inventory, the interaction is a failure and the Hearer updates its inventory by adding the transmitted word to it.
\end{itemize}

A unit of time ($t=1$) is defined as $N$~elementary steps (i.e., each agent is selected as a Speaker once in a unit of time, on average).
Motivated by the similarity to the Ising model, we report the results of the simplest (nontrivial) case with only two words, A and B, which can be communicated by agents. (It might be interesting, however, to consider also a multiple-word case.) The A-B version of the Naming Game, that was also examined in some other contexts \cite{baron-two-word,baron-two-word1}, has an obvious double-degenerate absorbing state, where all agents reach a consensus and retain in their inventories either only A or only B. Earlier simulations of various versions of the Naming Game inclined us to expect that our model will also evolve toward one of its absorbing states. 

\begin{figure}
\includegraphics[width=\columnwidth]{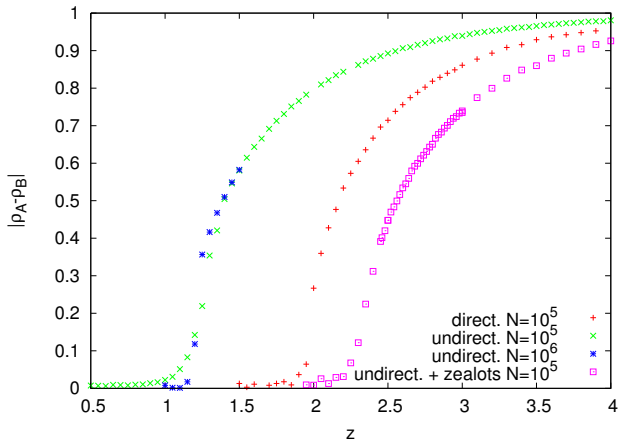}
	\caption{The absolute density difference $|\rho_{\textrm A}-\rho_{\textrm B}|$ as a function of the average degree $z$ in the A-B Naming Game on directed and undirected random graphs ($t_{\textrm relax}=10^5$; the average over 100 runs). For simulations on undirected random graphs with zealots (squares), the concentration of zealots was set as $\exp(-z)$ for each $z$, which corresponds to the concentration of sites with no out-links on directed random graphs.}
\label{naming-steady}
\end{figure}

Numerical simulations partially support our expectations (Fig.~\ref{naming-steady}).
	For $z>2$, we can see that the system evolves toward the broken-symmetry state, where one of the words prevails in the system. Let us notice that the examined values of $z$ are not so large and there is still a nonnegligible fraction of sites that do not belong to the giant cluster (which exists in the system above the percolation transition at $z=1$ \cite{dorog,newman,liplip2015}). In such separate clusters, a consensus is reached  independently, which thus has a negligible influence on the absolute density difference $|\rho_{\textrm A}-\rho_{\textrm B}|$, where $\rho_{\textrm A}$ and $\rho_{\textrm B}$ are the (normalized by the number of agents~$N$) densities of agents that finally have either the word A or B in their inventories.

Our simulations show that there is a range of $z$ ($1\sim 2$), where  a giant cluster forms, but the dynamics does not drive the system to the broken-symmetry state and keeps it disordered. 
For comparison, we also made simulations of the A-B Naming Game on an undirected random graph and the model seems to evolve to the broken-symmetry state for any $z>1$ (Fig.~\ref{naming-steady}). Let us notice that in this case the percolation threshold is also at $z=1$~\cite{albert}. 

The zero-temperature Ising model on an undirected random graph does not evolve to the broken-symmetry state but instead it gets trapped in a certain disordered state~\cite{castellano,svenson,haggstrom,liplip2015}. The Naming Game, similarly to the zero-temperature Ising model, has no bulk noise. It means that an agent that has only A in its inventory and is surrounded by agents which also have only A in their inventories cannot acquire B. However, contrary to the zero-temperature Ising model and similarly to the Voter model, there is an interfacial noise in the Naming Game. As a result, the interface between the A and B domains may fluctuate and increase its length (or even create some bubbles). In our opinion, due to such interfacial noise in the Naming Game, the frozen states do not form as they do in the zero-temperature Ising model, and the dynamics coarsens the system; finally, a broken-symmetry state is reached both in the undirected and (for $z>2$) directed versions of the model.

The above arguments explain the behaviour of the undirected version of the Naming Game. However, they do not clarify the behaviour of its directed version and, in particular, the nature of the transition around $z=2$. We propose an explanation, wich refers to the notion of zealots, i.e., those agents that never change their states. It is easy to realize that a site with no out-links will never be selected as a Hearer, and its inventory will remain unchanged (Fig.~\ref{graph}b). On a directed random graph, the probability that a randomly chosen site is a zealot equals $(1-z/N)^{N-1}$, where $z/N$ is the probability that there is a directed link from $i$ to $j$. In the limit $N\rightarrow \infty$, this probability equals $\exp(-z)$. Since a zealot remains unchanged, it creates in its vicinity a cloud of agents that are in the same state as the zealot. When $z$ is large, a concentration of zealots is small and such clouds  do not play an important role. As a result the system becomes ordered (except some clouds around zealots, which are in conflict with the bulk). However, when $z$ decreases, the concentration of zealots increases, and at a certain point (around $z=2$ for the Naming Game) cloud interference destroys the ordered state.

\begin{figure}
\includegraphics[width=\columnwidth]{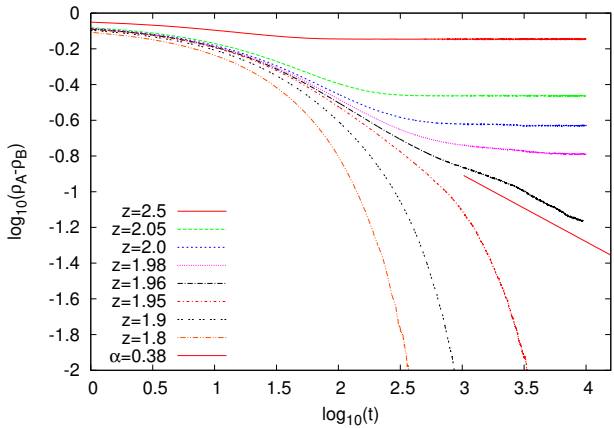}
\caption{The time dependence of the density difference $\rho_{\textrm A}-\rho_{\textrm B}$ in the A-B Naming Game on directed random graphs ($N=10^6$). Agents on sites with no out-links were set randomly as either A or B, and all the other were set to A. At the critical point ($z=1.96(2)$), we expect the power-law decay $\rho_{\textrm A}-\rho_{\textrm B} \sim t^{-\alpha}$ with $\alpha\approx 0.38(2)$.
Presented results are averages over 100 independent runs and only for $z$ close to the critical point averaging over 1000 runs was made.
}
\label{loglogng}
\end{figure}

	To confirm the above scenario, we performed simulations where initially all agents had only A in their inventories, except zealots, for which either A or B was chosen randomly. We calculated the time dependence of the density difference $\rho_{\textrm A}-\rho_{\textrm B}$ (Fig.~\ref{loglogng}). When $z$ is large, after an initial relaxation the difference $\rho_{\textrm A}-\rho_{\textrm B}$ saturates at a certain positive value.  When $z$ decreases, the asymptotic value of $\rho_{\textrm A}-\rho_{\textrm B}$ also decreases, but below a certain value the density difference $\rho_{\textrm A}-\rho_{\textrm B}$ seems to decay fast (perhaps exponentially fast) to~0. At $z=1.96$, the density difference most likely has a power-law decay, $\rho_{\textrm A}-\rho_{\textrm B} \sim t^{-\alpha}$, where $\alpha= 0.38(2)$.
In these simulations, the B-zealots provide the only perturbation to the rest of the system. For $z>1.96$, this perturbation is too small to destroy the A-dominance and $\rho_{\textrm A}-\rho_{\textrm B}$ remains positive. Below $z=1.96$, the B-zealots are sufficiently dense to destroy the A-dominance, and $\rho_{\textrm A}-\rho_{\textrm B}$ quickly decays to~0.
Let us notice that the dynamically determined location of the phase transition $z=1.96(2)$ (Fig.~\ref{loglogng}) agrees with the (visual) estimation based on the steady-state simulations, which start from a random initial configuration (Fig.~\ref{naming-steady}).
In our opinion, the dynamical method provides a much more accurate method to determine the location of this phase transition. Similar methods are  known to be very efficient for some models with absorbing states~\cite{hinrichsen,odor}.

The increasing density of zealots may destroy an ordering also in undirected models. We made simulations of the A-B Naming Game on undirected random graphs with a fraction $\exp(-z)$ of sites being set as zealots (i.e., agents which cannot be selected as Hearers, no matter what their coordination number is).
Numerical simulations show that, analogously to the directed graph, the transition takes place around $z=2.3$ (Fig.~\ref{naming-steady}). In certain models of opinion formation on undirected networks, the zealots were also shown to be responsible for a transition-like behaviour~\cite{szymanski,waagen,xie}.

\section{Ising Model}
As we have already mentioned, the present studies were motivated by our earlier work on the Ising model on directed graphs~\cite{liplip2015}. 
We analysed the heat-bath dynamics version, with a spin variable $s_i=\pm 1$ ($i=1,\ldots, N$) at each site of a network. The algorithm sequentially selects a site (say $i$) and sets $s_i=1$ with probability 
\begin{equation}
p(s_i=1)=\frac{1}{1+\exp(-2h_i/T)}, \ \ \ \ \ h_i=\sum_{k_i} s_{k_i},
\label{heat-bath}
\end{equation}
and  $s_i = -1$ is set with probability $1 - p(s_i=1)$.
The temperature-like parameter~$T$ controls the noise of the system and the summation
in Eq.~(\ref{heat-bath}) includes all sites~$k_i$ for which there is an out-link
from~$i$ to~$k$. Let us emphasize that we refer to the above system as the Ising model only because its dynamical rules bear some similarity to the heat-bath dynamics of the equilibrium Ising model. On undirected networks, such rules would correspond to the equilibrium Ising model, but for directed networks, we do not find such analogies \cite{sanchez,bray}. 

For the Ising model on  directed random graphs, sites with no out-links, according to Eq.~(\ref{heat-bath}), are set to~1 or~$-1$ with the same probability~$1/2$. In simulations, such flickers play a role similar  to randomly set zealots in the Naming Game. Analogously to the Naming Game, we performed dynamical simulations of the $T=0$ Ising model with the heat-bath dynamics Eq.~(\ref{heat-bath}), where initially all spins are set to~1 except for the flickers, which are randomly set to~$\pm 1$. 
In the limit $T=0$ Eq.(\ref{heat-bath}) implies that
\begin{equation}
p(s_i=1)= \left \{ \begin{array}{ll}
1 & \ {\rm for} \ h_i>0\\
0 & \ {\rm for} \ h_i<0\\
1/2 & \ {\rm for} \ h_i=0\\ 
\end{array}\right.
\end{equation} 

\begin{figure}
\includegraphics[width=\columnwidth]{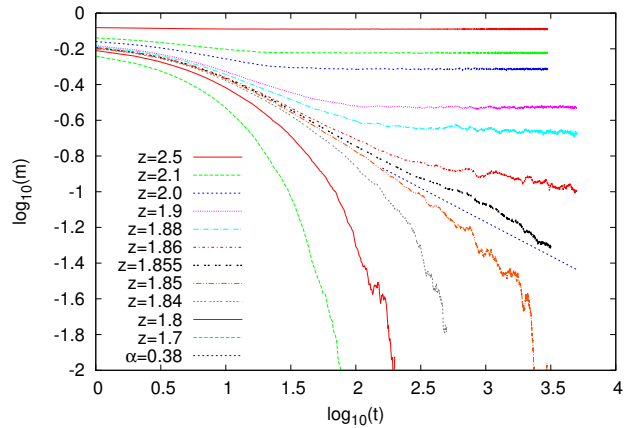}
\caption{The time dependence of the magnetization $m$ in the Ising model on directed random graphs ($N=10^6$). Spins on sites with no out-links were set randomly to~$\pm 1$, and all the other were set to~$+1$.  The results for each $z$ are averages over 100 independent runs. At the critical point ($z=1.855(5)$), we expect the power-law decay $m\sim t^{-\alpha}$ with $\alpha\approx 0.38(2)$.}
\label{loglogising}
\end{figure}

Numerical results (Fig.~\ref{loglogising}) show also a behaviour similar to that of the Naming Game. For large $z$, the  surface tension limits the influence of flickers and the system remains in the $m>0$ state. However, for decreasing $z$, the concentration of flickers increases; at a certain point it is sufficient to destroy an ordering and the system evolves toward $m=0$ states. The estimation of the transition point ($z=1.855$) is in a good agreement with the previous steady-state simulations as well as with the mean-field approximation~\cite{liplip2015}. Let us also notice that at the transition point the magnetization decay seems to behave similarly to the Naming Game, which suggests a certain universality of these transitions.

\begin{figure}
\includegraphics[width=6cm]{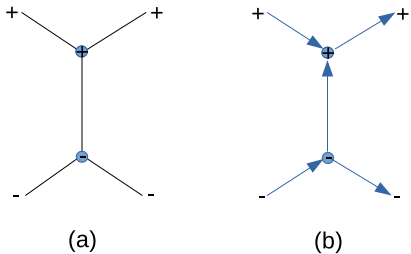}
\caption{(a) Boundary spins (encircled) of oppositely magnetized domains in the Ising model on an undirected network are frozen in their positions. (b) For the bottom boundary spin on a directed network, the contributions from its out-linked neighbours cancel out and the spin gets the values~1 or~$-1$ with equal probability~$1/2$.}
\label{blocking}
\end{figure}

What is, however, not quite well understood for us is the problem of why during coarsening the Ising model on a directed random graph reaches broken-symmetry ($m\neq 0$) states at all. Let us recall that on the undirected random graph such an evolution gets trapped in some disordered ($m=0$) configurations~\cite{castellano,svenson,haggstrom}. In the case of the Naming Game, we argued that it is the interfacial noise that precludes formation of some blocking configurations, and both in the undirected and directed versions, the coarsening ends up in the broken-symmetry ($\rho_{\textrm A}-\rho_{\textrm B} \neq 0$) states (Fig.~\ref{naming-steady}).
The absence of such  noise in the Ising model may trap the dynamics; a simple example is shown in Fig.~\ref{blocking}a.
We can only speculate that on directed random graphs the formation of blocking configurations is less likely. Indeed, for the orientation of bonds shown in Fig.~\ref{blocking}b, the spin at a boundary of the negative domain takes $\pm 1$~values with equal probabilites~$1/2$. Of course, some other orientations of bonds might be blocking, but large domains are, in our opinion, unlikely to have only blocking-type boundaries. Any configuration of the type shown in Fig.~\ref{blocking}b would create a passage for the coarsening to proceed. 

\begin{figure}
\includegraphics[width=6cm]{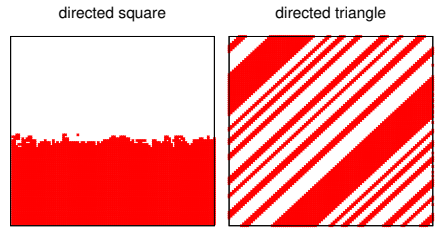}
\caption{(Left) The directed square-lattice Ising model with the heat-bath dynamics Eq.~(\ref{heat-bath}) at $T=0$. Starting from the initial configuration with a flat interface,  the system demonstrates the presence of an interfacial noise already after $t=10$ steps. (Right) Coarsening on the directed triangular lattice leads to a frozen stripe-like pattern.}
\label{configsq}
\end{figure}

The above reasoning suggests that boundaries in the directed Ising model may actually fluctuate, which would make the model similar to the Naming Game. An additional evidence of such interfacial noise in directed networks comes from the visual inspection of configurations of the directed square-lattice Ising model. In this model, the out-links are vertically oriented, e.g., in North and East directions \cite{bray,liplip2015}. The model is known to be disordered for any positive temperature but has a ferromagnetic ground state.  Starting from the configuration with a flat interface (the same as for models in Fig.~\ref{config}), the zero-temperature heat-bath dynamics roughens the interface already after $t=10$, which clearly shows the existence of an interfacial noise (Fig.~\ref{configsq}). It is easy to show  that in this model at $T=0$, the ground-state configurations are the only configurations with all spins fixed. Any other configuration must contain some free spins, which will sooner or later bring the model to the ground state. In particular, any interface must thus contain some free spins, which we identify as an interfacial noise.

In the undirected (ordinary) Ising model (at $T=0$),  configurations with the a flat interface (as in Fig.~\ref{config}) do not contain free spins and thus there is no interfacial noise. However, some models on directed networks have also this property. For example, in the Ising model on a triangular lattice (North, East, North-East), a stripe-like pattern develops during coarsening from a random initial configuration, and it is also easy to show that it contains only frozen spins (Fig.~\ref{configsq}).

Thus, we can see that on regular directed lattices there are Ising models both with a presence or absence of the interfacial noise. Such behaviour does not resolve the problem of directed random graphs but, in our opinion, it may suggest that some interfacial noise might indeed be present in such systems, which is at least sufficient to bring the Ising model to the broken-symmetry state (not necessarily fully magnetized). We admit, however, that some more convincing  arguments are needed to explain why coarsening in the Ising model on directed graphs leads to the broken-symmetry states.


\section{Voter Model}
Yet another model of the agreement dynamics is the Voter model. In a directed two-state version ($s_i=\pm 1$) of this model, a randomly selected spin takes the value of one of its randomly selected out-neighbours. A site that has no out-links is considered a zealot and its state is fixed (a modification with such sites being  flickers would exhibit a similar behaviour).

\begin{figure}
\includegraphics[width=\columnwidth]{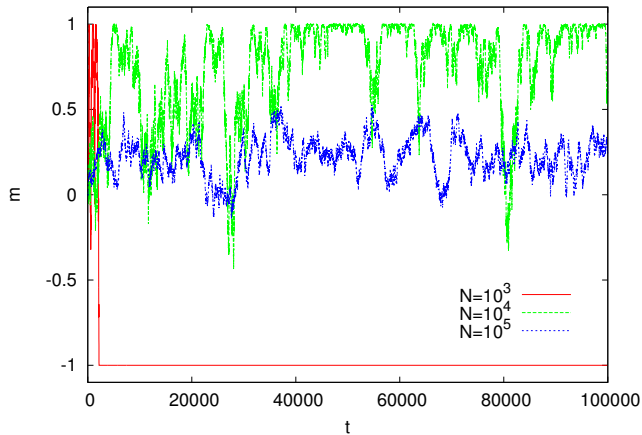}
\caption{The time dependence of magnetization $m$ in the Voter model on  directed random graphs with $z=7$ and random initial configurations. For small systems ($N=10^3$),  consensus ($m=\pm 1$) is easily reached. For larger systems, consensus is not reached due to the presence of zealots.}
\label{time7voter}
\end{figure}

On undirected random graphs, the Voter model is known to order, albeit on a time scale diverging with the system size~$N$~\cite{sood}. Our simulations show that on  directed random graphs, the model seems to order for small system size~$N$, however, for larger~$N$, it remains disordered. Such  behaviour appears for any~$z$ and is rather easy to understand. Let us notice that contary to the Ising and Naming Game models, the Voter model has no surface tension. It means that a cloud around a zealot can freely spread throughout the system. If there are at least two zealots of opposite opinions in the system, a broken-symmetry ($m\neq 0$) asymptotic state will never be reached. Since the average number of zealots in the system equals $N\exp(-z)$, we obtain that $N\approx \exp(z)$ sets such a size scale, below which most likely there are no zealots and the system orders. Our simulations confirm these considerations. For $z=7$ ($e^7 \approx 1100$) and $N=10^3$, the system quickly reaches  full consensus (Fig.~\ref{time7voter}).
For larger $N$, however, zealots prevent ordering. When $N$ is not too large ($N=10^4$), fluctuations in the number of zealots or in their distribution may keep the system in the broken-symmetry state, but for larger $N$ ($N=10^5$), the symmetry seems to be restored (Fig.~\ref{time7voter}). 
Let us also mention that the role of the zealotry was already examined in the some other versions of Voter model \cite{mobil2007,masuda}.

\section{Conclusions}
In the present paper, we analyzed some agreement-dynamics models placed on directed random graphs. 
A fraction of sites of such graphs  have no out-links and in the context of the agreement dynamics, these sites become either zealots or flickers.  We have shown that these sites play a very important role. In the Voter model, the absence of a surface tension implies that the zealots  will freely affect the system, thus  preventing reaching consensus. In the Naming Game, the influence of zealots is limited by the effective surface tension. However, when the average degree~$z$ decreases, the concentration of zealots increases and at a certain threshold the ability of the system to reach the consenus is lost. A similar behaviour was found in the zero-temperature Ising model and some power-law characteristics at the threshold value suggest that the Ising model and the Naming Game exhibit some kind of universality. Reaching the consensus in the Naming Game is not surprising since the model exhibits the so-called interfacial noise, which precludes formation of blockades that could trap the coarsening dynamics at some disordered configurations. Less obvious is reaching the consensus in the $T=0$ Ising model. The experience we gained in the case of undirected graphs or regular lattices suggests that the model does not have the interfacial noise and, indeed, on undirected random graphs (but also on some regular lattices~\cite{lip2000}), the coarsening  dynamics gets trapped in some disordered configurations. The situation is, however, more subtle on directed graphs. Prompted by the example of a directed square lattice, we suggested that the Ising model on a directed random graph may actually exhibit some interfacial noise and that is why the coarsening does not get trapped. Qualitatively, one might consider that both in the Naming Game and in the Ising model the density of zealots (that is controlled by the average degree $z$) plays the role of a temperature-like parameter, which controls the level of noise in the system. At large~$z$ the density of zealots is small and a weak noise keeps the system in a broken-symmetry (consensus) phase. For sufficiently small~$z$ a large density of zealots and thus strong noise will bring the system to the disordered state.   

It would be perhaps interesting to extend our analysis to some other directed networks. As for the Ising model, some results were reported only for the Barab\'asi-Albert networks, showing that the behaviour of such systems depends to some extent on the chosen dynamics~\cite{lima}. It would be also interesting to examine some modifications of our models where concentration of zealots (i.e., of sites without out-links) would be different than in Poissonian random graphs. Moreover, the broken-symmetry (biased) case  where a certain opinion is represented by most (or all) zealots might lead to a different behaviour. Analysis of directed versions of some other agreement-dynamics models~\cite{axelrod,sznajd,weisbuch}, as well as placing them on networks that would be more appropriate to describe social relations would be also desirable.

Acknowledgements: The research for this work was supported by Narodowe Centrum Nauki (NCN, Poland) Grant No. 2013/09/B/ST6/02277 (A.L.).

\end {document}